# Evaluating feasibility of batteries for second-life applications using machine learning

Aki Takahashi[1], Anirudh Allam[1], Simona Onori[1,2,*]

[1] Department of Energy Science and Engineering, Stanford University, Stanford, CA 94305, USA

**Summary**

This paper presents a combination of machine learning techniques to enable prompt evaluation of retired electric vehicle batteries as to either retain those batteries for a second-life application and extend their operation beyond the original and first intent or send them to recycle facilities. The proposed algorithm generates features from available battery current and voltage measurements with simple statistics, selects and ranks the features using correlation analysis, and employs Gaussian Process Regression enhanced with bagging. This approach is validated over publicly available aging datasets of more than 200 cells with slow and fast charging, with different cathode chemistries, and for diverse operating conditions. Promising results are observed based on multiple training-test partitions, wherein the mean of Root Mean Squared Percent Error and Mean Percent Error performance errors are found to be less than 1.48% and 1.29%, respectively, in the worst-case scenarios.

**Introduction**

Lithium-ion battery technology is used in a wide variety of industries due to superior energy and power density among available electrochemical devices, as well as their cycling capability. The loss in performance observed over time is due to multiple degradation phenomena, including loss of active material, corrosion, passivation, lithium plating, and solid electrolyte interphase layer growth.[1]

Upon retirement from their first life application, batteries are sent to warehouses where they are piled up and stored waiting to be screened. Their health history is unknown and therefore it is critical to be able to assess the level of deterioration to decide whether the battery can be safely utilized in later applications such as backup power, residential storage, EV charging, and utility scale storage[2] to capture the maximum value -both economic and environmental, or sent to recycling facilities. Applications span from According to a McKinsey report, the supply of second-life Electric Vehicle (EV) batteries could surpass 200 gigawatt -hours per year by 2030, with the potential to meet half of the forecast global demand for utility-scale energy storage in that year.[3]

---

[2] Lead Contact
[*] Corresponding author: sonori@stanford.edu

Battery state of health (SOH) is a metric for diagnosing battery degradation during testing and operation. While many unique invasive measurements to diagnose the health of the cell are possible during the testing phase in a laboratory setting, instead during real-time operation, only non-invasive measurements in the form of temperature, voltage and current are accessible for diagnostic purposes, which are typically done via model-based or data-driven approaches.Recent research work has shown promising results in assessing battery health, in terms of impedance, capacity, or power, via data-driven approaches.[4,5,6] For example, neural networks models independently find relationships between degradation indicators and battery health,[7,8] in support vector regression and random forest algorithms differences between features are defined in order to learn the relationship between features and the response.[9,10] Gaussian Process Regression (GPR), a nonparametric approach to regression, has become an increasingly popular method due to its interpretable outputs and the available prediction uncertainty.[4,11,12] There are different approaches to GPR, namely, where training is done on some early cycles and predictions are made thereafter,[13,14] or where models are trained on some cells and then predictions are made on different test cells.[12,15] The former approach assumes that the battery undergoes similar use cases in both early and later cycles, making it an inappropriate approach for repurposing battery applications. On the other hand, with the latter approach the prediction algorithm can be extrapolated to cells with similar operating conditions. Currently, there is lack of a universal approach to battery health assessment and life prediction, across chemistries, due to the nonlinear nature of battery degradation dynamics.[5,16,17] In addition, scalability of data-driven models must be addressed to prepare for the influx of battery data.[18]

In this paper, a data-driven approach which combines ensemble methods with GPR is proposed to quickly estimate overall battery capacity. The model uses features from voltage and current information in a limited window of the charge profile. Time is not used explicitly, which, while valuable[13] may not be practical when the battery undergoes incomplete charging or discharging processes. In our approach, we use bagging for the ensemble learning. This work highlights the importance of selecting statistical features over a time horizon that embeds degradation information to develop robust data-driven models that accurately and quickly assess battery health.

Contributions of this work are as follows: 1) an ensemble based GPR model is proposed to estimate overall battery health – in terms of capacity loss - quickly and efficiently, 2) features from voltage and current measurements are proposed for fast screening, 3) the validation of the model is tested on multiple datasets of lithium-ion batteries, each with different experimentation and aging processes.

**Results**

**Battery health indicators (HI)**

Health indicators to describe the battery's lifetime performance include, $SOH_C$ and $SOH_E$, defined as a ratio of the battery's actual capacity and actual energy (at any given point in time) relative to the nominal capacity ($Q_{nom}$) and energy ($E_{nom}$), respectively, measured at the beginning of life, as follows:

$$SOH_C(t) = \frac{Q(t)}{Q_{nom}} \cdot 100\% \tag{1}$$

$$SOH_E(t) = \frac{E(t)}{E_{nom}} \cdot 100\%, \tag{2}$$

where $Q(t)$ and $E(t)$ are the capacity and energy at time *t*. While a deterioration of 20% in $SOH_C$ is the industry standard for battery end of life, $SOH_E$ is supplemented with changes in the battery voltage. As the battery is used, internal resistance increases due to the aging mechanisms, resulting in a shift of the capacity-voltage curve over time.[19] Thus, $SOH_E$ deterioration can also be used to identify the increase in resistance of the cell.

**Data Processing**

Datasets from publicly available repositories, summarized in Table 1, were used in this work (see Experimental Procedures) upon proper processing. Despite the different operating conditions, chemistry, and aging trajectories of the batteries as noticed in Figs. 1.a, 1.b, and 1.c, a strong linear correlation among $SOH_C$ and $SOH_E$ is observed in Figs. 1.d, 1.e, and 1.f for NMC, LCO, and LFP, respectively. Therefore, the analysis that follows is conducted using the capacity-based HI, $SOH_C$, as aging metric, denoted as SOH henceforth.

| Dataset | Cathode | Form factor | Nominal Capacity [Ah] | Temp-erature [°C] | Aging protocol | Charging C-rate [1/h] | Voltage Range [V] | Number of cells |
|---|---|---|---|---|---|---|---|---|
| [16] | NMC | Pouch | 0.74 | 40 | Artemis driving schedule | 1 | 2.7 – 4.2 | 8 |
| [17] | LCO | 18650 | 2.1 | Room | Statistically random discharge | 1 | 3.2 – 4.2 | 20 |
| [5] | LFP | 18650 | 1.1 | 30 | 2-step CC-CV charge, CC discharge | 1 – 6 | 2 – 3.6 | 124 |
| [20] | LFP | 18650 | 1.1 | 30 | 4-step CC-CV charge, CC discharge | 4 – 8 | 2 – 3.6 | 45 |

**Table 1**: Summary of different datasets used in this work and their experimental conditions

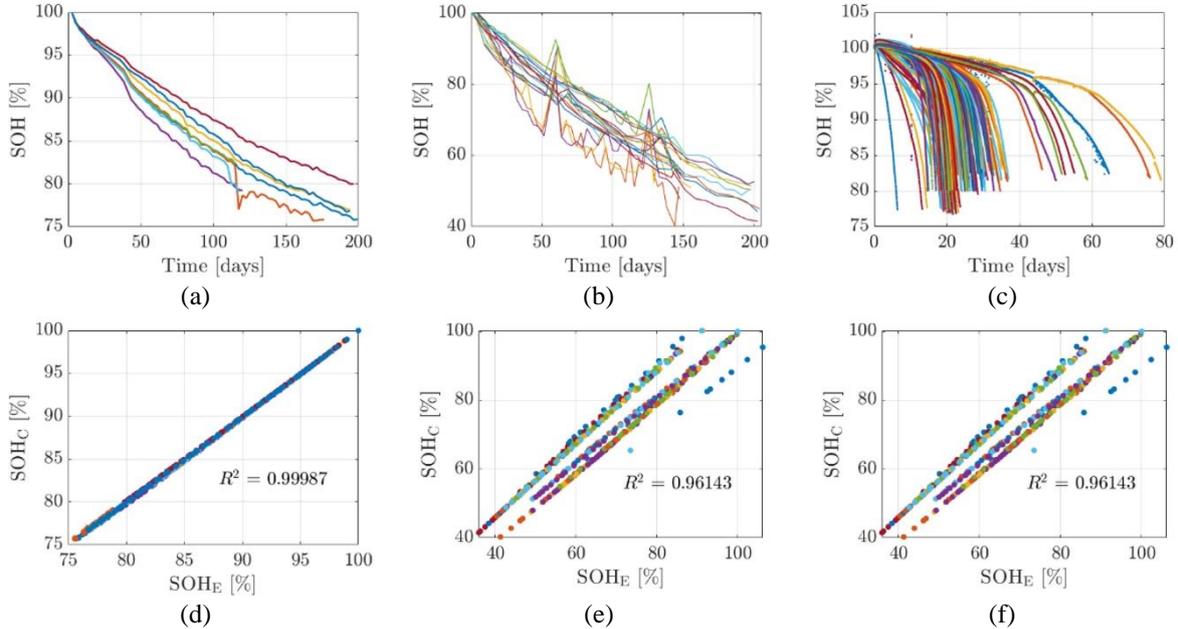

**Figure 1**: SOH vs. time for different cell chemistry that operate on a different range of days as well as aging procedures (a) NMC,[16] (b) LCO,[17] and (c) LFP.[5,20] Correlation and change in $SOH_C$ and $SOH_E$ for all cells for (d) NMC, (e) LCO, (f) LFP. A linear relationship can generally be observed for all three groups of cells, with a high correlation coefficient for all three datasets, irrespective of the testing operating conditions. In (d), the linear fit is almost perfect, which suggests that $SOH_C$ and $SOH_E$ are interchangeable for this chemistry. In (e), correlation is high, albeit lower than the other two chemistries. For (f) it should be noted that in [5], the capacity-voltage feature used to predict capacity degradation was a measure of $SOH_E$, which can be observed to decrease faster compared to $SOH_C$. However, $SOH_E$ was not observed to deteriorate faster than $SOH_C$ in the other two datasets.

Feature generation is carried out on the charge cycles since constant current and/or constant voltage (CC-CV) data is generally standardized as opposed to the discharge cycle, which is heavily determined by the intended application and user-dependent. It is observed that the CC and CV regimes are key regions of the battery operation where deterioration is highly perceptible, and therefore extracted to generate features. To that end, for the CC and CV processes, due to the diverse electrode chemistry and aging conditions of the datasets, observations from the certain pre-selected regions of the voltage and current data utilized. For the LFP cells, 30 seconds worth of lower portion of voltage data during CC and upper portion of current data during CV are considered.; Data spanning 3.65V to 4.2V during CC for 25 minutes or less (as low as 70 seconds depending on the SOH of the cell) and the whole current profile worth two to three hours during the CV phase are considered for LCO cells. Finally, one-hour worth of voltage data during the CC phase and no CV phase are considered for NMC cells. Since the datasets in Table 1 are characterized by different charging rates, the measurements used for generating features during the CC-CV phase are sampled or selected accordingly to maintain consistency. The sampling rate for the features is an important design parameter as excessive data collection can result in a computationally prohibitive model. For higher C-rates, the battery reaches the maximum voltage sooner, so a limited number of measurements are available to sample. However, for lower C-rates, it is not necessary to sample measurements frequently since the change in the voltage and current curves occur slowly, and high frequency sampling would lead to a large amount of data for the model to train on. The NMC and LCO datasets have >1 hour charge durations with low current. For these datasets, we take either voltage or current measurements every 10 seconds to get sufficient data until end of CC or CV. As for the LFP dataset, where the C-rates are much higher, we take measurements every two seconds in a 30 second window for CC and 60 seconds for CV to capture. This also allows us to demonstrate the capability of this approach in both slow and fast charge regimes, as well as a limited window of measurements. Fig. 2 illustrates the voltage vs. time and current vs. time curves over many CC-CV cycles of a representative cell in the LFP dataset.[5,20] During the CV portion of the charge profiles at the known voltage, current signals are also analysed for feature extraction. Note that for the NMC dataset,[16] the cells did not undergo CV, hence CV is not considered in this case.

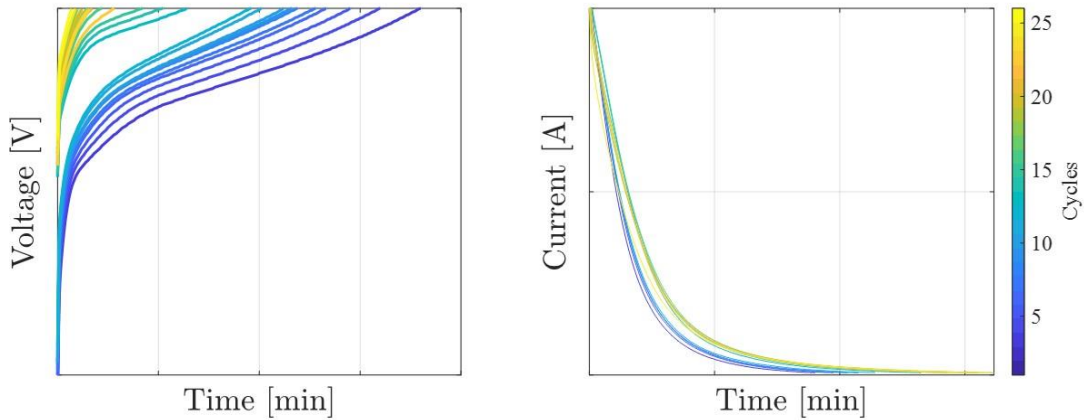

**Figure 2**: Voltage curve during CC (left) and current curve during CV (right) for a sample cell from the LCO dataset[17] for feature generation.

**Feature Generation and Selection**

For each of the CC and CV pair of datasets, 7 simple statistical measures are used to characterize the data, which are summarized in Fig. 3.a. The statistics are applied to voltage during CC and current during CV, respectively. The main benefit of using said statistics is their simplicity in calculation during operation. In addition, they do not rely on derivatives, as instead seen in incremental capacity and differential voltage analysis methods, which can introduce noise without using a slow characterization procedure.[21] Lastly, the features are shifted by subtracting the same metric calculated at the start of life.

Once the features are generated, a feature selection technique is used to eliminate redundant and noisy features. This reduces the dimensionality of the model and allows for efficient and accurate estimation of SOH. There are a variety of selection techniques, such as filtering, wrapping, and fusion, with each varying in purpose and complexity.[22] Here, we use the fastest filter method via correlation coefficient, such as Pearson or Spearman, to quickly determine which features correlate best with the predicted variable.

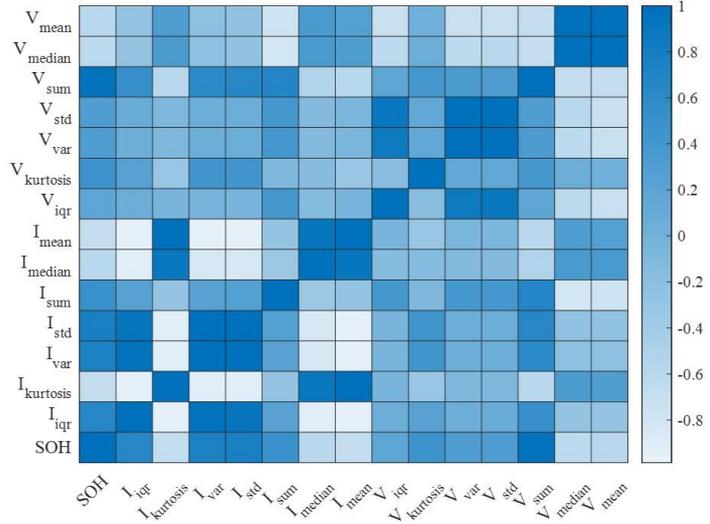

| Statistics | Meaning |
|---|---|
| Mean | Average over recorded values |
| Median | Central value of distribution |
| Sum | Adding all recorded values |
| Standard Deviation | Measure of spread with respect to mean |
| Variance | Measure of spread |
| Kurtosis | Measure of deviation from normal distribution at extreme values |
| Interquartile range | Measure of spread between 25% and 75% percentile |

(a)          (b)

**Figure 2**: (a) List of statistics applied on the voltage and current measurements. The selected statistics are chosen for their ease of calculation, and their definitions are briefly explained. (b) Heatmap displaying the relationship between the 14 features and SOH for a sample subset of 120 cells from the LCO dataset. A value for Spearman coefficient $\rho$ close to 1 represents a stronger correlation, or similarity, between the two metrics. Out of these candidates, the best combination of features is selected based on their uniqueness and strong predictive capability.

The Pearson correlation coefficient assumes a Gaussian distribution for the vectors and a linear relationship between two variables. This assumption is sometimes not true due to the nonlinear nature of capacity degradation. Hence, the Spearman coefficient is found to be more suitable and therefore used. The equation is as follows:

$$\rho = 1 - \frac{6 \sum_{i=1} d_i^2}{N(N^2-1)} \quad (3)$$

where $\rho$ Spearman's rank correlation coefficient, $i$ is the $i$-th value in the corresponding feature vector, $d$ is the difference between the two sorted ranks of features and SOH, and $N$ is the number of data points in the dataset. This expression looks for a monotonic relationship between features and responses, meaning the magnitude of $\rho$ increases as the response either increases or decreases as the feature increases. Once the magnitude of $\rho$ was found for all features, we select 10 features with highest value, while also looking for dissimilar features ($|\rho| < 0.8$ between features), as done in.[15] An illustration is shown in Fig. 3.b for the LFP dataset. An example of the relationship between features and SOH for representative cells is shown in Fig. S1, S2, and S3 for NMC, LCO, and LFP cells, respectively, in the Supplemental Information section.

**Scalable Gaussian Process Regression**

GPR is a nonparametric probabilistic algorithm used to make predictions based on Gaussian distributions of features and response.[23] Specifically

$$GPR\ (\mu(x), k(x, x')), \qquad (4)$$

is a Gaussian Process (GP) with mean function $\mu(x)$ and covariance function $k(x, x')$, where $x$ and $x'$ are training inputs. If each set of the inputs has a joint Gaussian distribution, then the whole set of inputs forms a joint Gaussian distribution. This means that the predicted response has a distribution, with $M$ and $k$ defining the response and its uncertainty. Further details are provided in the Experimental Procedures section.

While GPR models are effective, they suffer from lack of scalability, since for model training there are $O(n^3)$ and $O(n^2)$ computation and storage requirements (in big-$O$ notation), while the cost of a single prediction is $O(n^3)$, where $n$ is the number of datapoints being trained on. This is a significant problem for second-life battery health assessment and other practical applications where large number of batteries need to be processed[18] in a short period of time. To mitigate this, one approach is to use bagging (bootstrap aggregating), an ensemble learning technique used mainly for random forest algorithms but transferable to other machine learning algorithms, including GPR.[24] From a large dataset $m$, different bags of size $n$ are created by randomly sampling from the dataset with replacement, meaning some examples can be selected more than once. Once the $m$ sampled datasets are created, a GPR model is trained on each dataset to create $m$ models. For subsequent predictions then, these models are combined using some aggregating technique, such as a simple average or weighted average of the predicted output. It should be noted that this is different from Bayesian Committee Machines, which instead uses scalable GPs, where all of the training data is split into subsets instead of random sampling with replacement.[25]

Bagging is attractive for its reduction of variance since it creates multiple models instead of relying on a single model, which means if one of the models is overfitting the training data, then the other models can work together to improve the prediction accuracy of the test dataset. It also allows for parallel computing since each of the bagged models undergo training and prediction separately. Each of the GPR models can make their own predictions, and when merged, their net combined performance improves noticeably.[24] We use a weighted average to account for the predictive capability of each of the bags (see Experimental Procedures section).

The computational time needed to train the bagged model can be reduced significantly by using bags smaller than the whole dataset, as well as using parallel computing features. In general, the bagged approach reduces the amount of training time since there is no need to perform optimization on a large feature set. With bagging, computational burden becomes $O(mn^3)$, meaning that with smaller selection of $m$ and $n$ GPRs can be simplified exponentially while still achieving strong accuracy.[26] Specifically, the data used by the models can be compared using the following factor reduction of data (FRD) metric:

$$\text{FRD} = \frac{N}{m \cdot n} \quad (5)$$

where $N$ is the number of datapoints in the original data set. For constant $N$, the FRD can be increased by changing $m$ and $n$. A larger FRD leads to a significant increase in model speed for training and prediction when combined with parallel computing.

**Model Performance**

To measure model performance, an approximate 70-30 training-test split is randomly created based on individual cells. In the context of battery aging with data collected over a long period of time, the model is trained on the entire aging trajectory of 70% of cells and tested on the remaining 30% of cells. In addition, a GPR model without bagging, referred to as the baseline model, is developed to highlight the performance improvement of the proposed GPR with bagging approach.

Feature selection is conducted on the training set and features with low correlation are eliminated, as observed in the heatmap in Fig. 3.b. Then, the bagged models are created and trained. Finally, the models perform capacity assessments on all available test data, and the results are aggregated using an averaging rule. An example of the assessment process is shown in Fig. 4. This process is repeated several times for smaller datasets to simulate a greater number of tested cells. To evaluate model performance, we use Root Mean Square Percent Error (RMSPE) and Mean Percent Error (MPE). This combination allows to report model performance based on the presence of outliers and overall model performance. Formal definitions of RMSPE and MPE are described in the Experimental Procedures section.

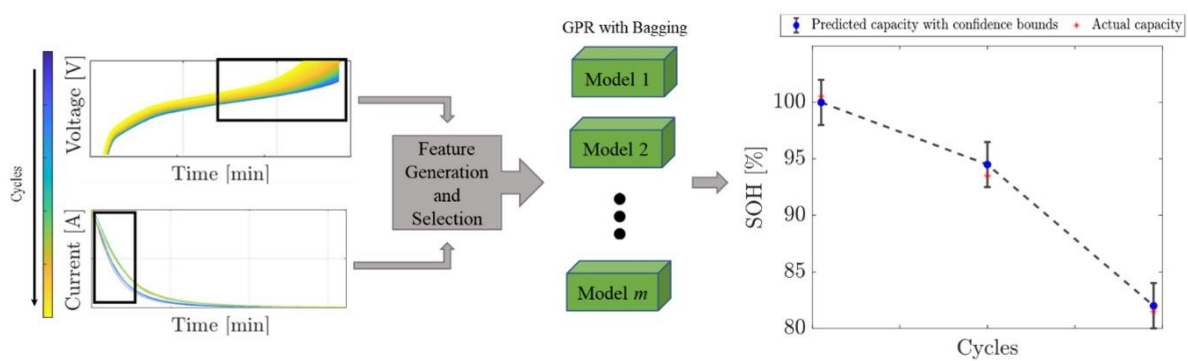

**Figure 3**: Graphical representation of fast ensemble-based SOH estimation framework based on GPR with bagging. Current and voltage data collected over a single charging curve are extracted from a prespecified window and processed to generate features. The features are passed through the bagged models to generate a SOH estimation with uncertainty bounds, which over time generates a capacity fade envelope.

**SOH Estimation Results**

For all datasets, $m$ was initially selected to be half of the number of training cells, with sample size $n$ being close to the average number of characterization cycles for each cell, then modifications were made based on performance. The selected values of $m$ and $n$ for each chemistry is shown in Table 2.

The distribution of SOH error for all individual training and test datapoints are shown in Figs. 5.a, 5.b, 5.c, and Figs. 5.d, 5.e compiles the MPE and RMSPE of individual cells into boxplots, Table 3 lists the model performance metrics in terms of median and mean of RMSPE and MPE Boxplots in Fig. 5 are used instead of conventional histograms to demonstrate performance on individual cells in addition to compare the different chemistries. Error mean and median are reported to describe the distribution, in comparison to the baseline in Table 3. Average training and prediction time for the 70-30 splits are measured using a i7-9750H CPU @ 2.60GHz and the parallel computing feature in MATLAB with 4 workers.

| Chemistry | $m$ (number of models) | $n$ (sample size) | FRD (factor reduction) |
|---|---|---|---|
| NMC | 3 | 20 | 8 |
| LCO | 7 | 30 | 2 |
| LFP | 20 | 200 | 23 |

**Table 2**: Values of $m$ and $n$ used in model implementation for each cathode chemistry. FRD in training data is found by dividing the typical amount of training data by the number of examples used in the bagged approach. This leads to an exponential decrease in training time.

| Metric | NMC | | LCO | | LFP | |
|---|---|---|---|---|---|---|
| | **Bagged** | **Baseline** | **Bagged** | **Baseline** | **Bagged** | **Baseline** |
| RMSPE$_{median}$ (%) | **0.3384** | 0.35132 | **1.099** | 1.109 | **1.266** | 1.2922 |
| RMSPE$_{mean}$ (%) | **0.2464** | 0.39763 | **1.277** | 1.5409 | 1.475 | **1.4624** |
| MPE$_{median}$ (%) | 0.286 | **0.26088** | 0.839 | 0.88566 | **0.907** | 0.93055 |
| MPE$_{mean}$ (%) | **0.28486** | 0.29896 | **0.925** | 1.194 | 1.286 | **1.0958** |
| Training Time (s) | **0.024061** | 2.0465 | **0.1613** | 3.1871 | **4.1123** | 123.8961 |
| Prediction Time (s) | **0.021394** | 0.067256 | 0.031035 | **0.028035** | **4.2562** | 4.2789 |

**Table 3**: Comparison of the median and mean of RMSPE and MPE, and training/prediction times for all three datasets between baseline and bagged. All datasets are tested to have a total of 300 test cells based on randomized bags. Training time and prediction time are reported as averages of multiple 70-30 splits of the data.

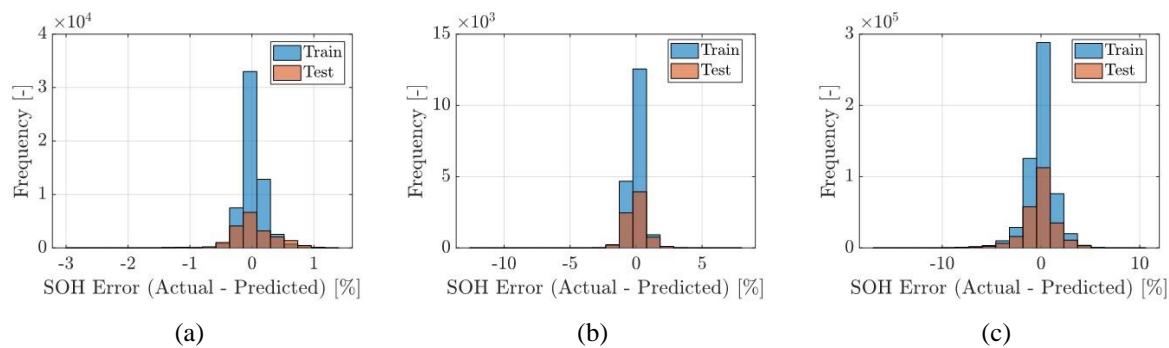

(a)　　　　　　　　　　　　(b)　　　　　　　　　　　　(c)

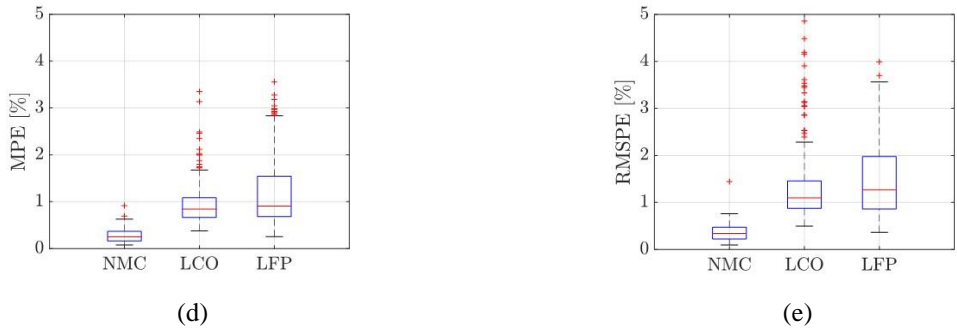

**Figure 4**: Distribution of SOH error with 20 bins for all individual training and test datapoints for (a) NMC, (b) LCO, (c) LFP cells. Boxplots show the median, 25th, and 75th percentiles, and the outliers (points past the interquartile range) for the MPE (d) and RMSPE (e) of 300 iterations of test cells.

## LCO Dataset

14 cells are selected for training and 6 were used for testing in each iteration of assessment. Assessments are made to the end of available data (during CC phase: 3.65V to end of CC, and for CV phase: full curve). We select 7 bags with sample size 30 to reduce the computational time to train the algorithm. 300 assessments are made based on 50 iterations. The median RMSPE and MPE are below 1.5% as shown in Figs 5.a and 5.b. Training time reduction of 2 orders was also found for this dataset. In this dataset the sum of the voltage values was an important feature as shown in Fig. S1 in the Supplemental Information section. This may be due to the long charge characterization, in that the CC regime was significantly longer and hence the feature generation captured more voltage points to add up in early cycles of the battery. This dataset shows the capability of SOH estimation far past the standard end of life of 80%, as data is available until 50% SOH.

## NMC Dataset

This dataset has the least cells; however, the 70-30 split still turns out to generate valid results with 6 cells used for training and 2 cells for testing in each iteration. 3 bags were created, each with 20 sample points, which reduced the training data to approximately 1/8 of its original size for each iteration, resulting in an exponentially faster solution. Since there are not many cells the process is repeated 150 times to get a total of 300 estimates. The results in Table 3 show that the bagged GPR estimation is more accurate than the baseline. Table 3 also reports that there was a training time reduction of 2 orders of magnitude compared to the baseline model.

Some of the higher RMSPE errors are likely due to the sudden drop in SOH past 50% in one of the cells; MPE is significantly lower than the other datasets since the effect of this single anomaly is small. The low error can also be attributed to the high correlation coefficient (>.95) between the CC features and SOH for cells, which is shown in Fig. S2 in the Supplemental Information section. It should be reiterated that these cells do not undergo a CV regime during the charging process, meaning that while we look for the 10 best features there are only 7 features used.

**LFP Dataset**

Since this dataset has the most data, the benefits of implementing an ensemble-based SOH estimation framework based on GPR with bagging are observed from reducing the computational time without sacrificing significant accuracy. The baseline model was impossible to store in memory with the original training data, so 2000 datapoints were instead randomly sampled out of about 90000 datapoints for each training iteration. Four cells were also anomalous in measurements, so they were excluded from analysis, leaving 165 cells. A training test split uses 115 cells for training and 50 cells for testing. Training data for the bagged models was reduced to less than 1/20 of a typical training set size. There is a training time reduction of 2 orders of magnitude compared to the baseline model, as shown in Table 3. For each test set there are 50 cells, so with 6 iterations there are 300 assessments made, whose error is shown in Fig. 5.d and 5.e. The median MPE was under 1%, which means that for all the three chemistries the ML model was able to achieve under 1% MPE based on the median. The error is slightly higher compared to the other two datasets, which is to be expected given the wide variety of fast charging protocols and large size of the dataset.

**Effect of $m$ and $n$**

In bagging techniques, there is a convention to choose $n$ to be the same size as the training data whereas $m$ is arbitrarily selected.[27] However, for GPs it is impractical to use large datasets due to the increasing training time and storage space, as well as the diminishing improvements in accuracy.[26] Hence, we analyze the effect of changing $m$ and $n$ in a set domain to see how much improvement in speed is achieved. First, the sample size $n$ is selected, then $M$ bags are created, where $M$ is the greatest number of bags used. Then, we train all $M$ models and use small batches of size $m$ out of the $M$ models for assessment. This methodology allows for comparisons between performance of bagged models with the same $n$ but different $m$, as the same subsamples are used and performance is measured largely based on $m$, not random sampling. The process is repeated 10 times for each dataset.

Since the datasets are of different sizes, the number of bags and sample size are different for each chemistry, which is outlined in Table 4. Beyond the values selected, we notice diminishing changes in performance. The results for MPE and RMSPE for different combinations of $m$ and $n$ are observed in Fig. 6.

In general, it can be observed in Fig. 6 that increasing $m$ and $n$ decreases the average MPE and RMSPE. Interestingly, the NMC dataset, which only contains 8 cells, had worse estimation with larger bags, albeit the increase in error is not significant. This may be due to the high predictive capability of the CC features and thus, larger number of bags results in making estimations more overconfidently. Increasing the number of bags past 10 did not have significant improvements, meaning that it is possible to make good SOH assessments with as little as 10 sample points each in 10 bags, which reduces the amount of training data by a factor of 4 and exponentially decreases computational time.

| Dataset | $m$ | $n$ |
|---|---|---|
| NMC | 2 to 100 | 10 to 150 |
| LCO | 2 to 50 | 10 to 150 |
| LFP | 2 to 80 | 50 to 1000 |

**Table 4**: Range of $m$ and $n$ used for the different datasets.

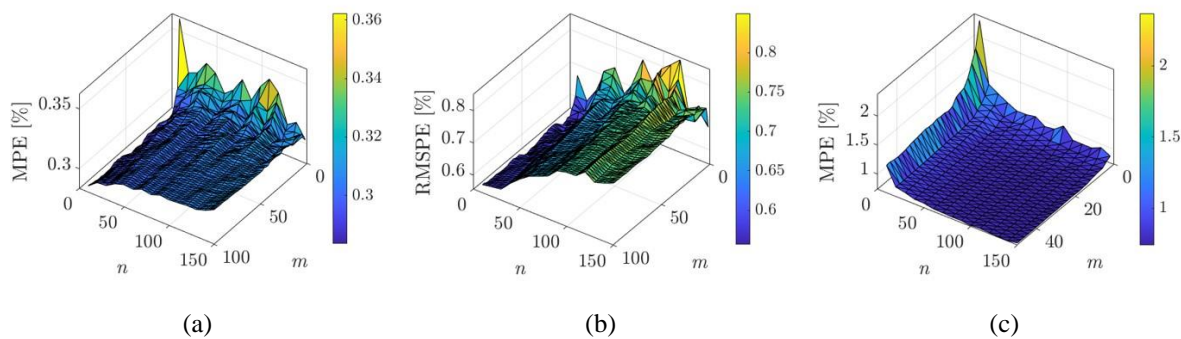

(a)        (b)        (c)

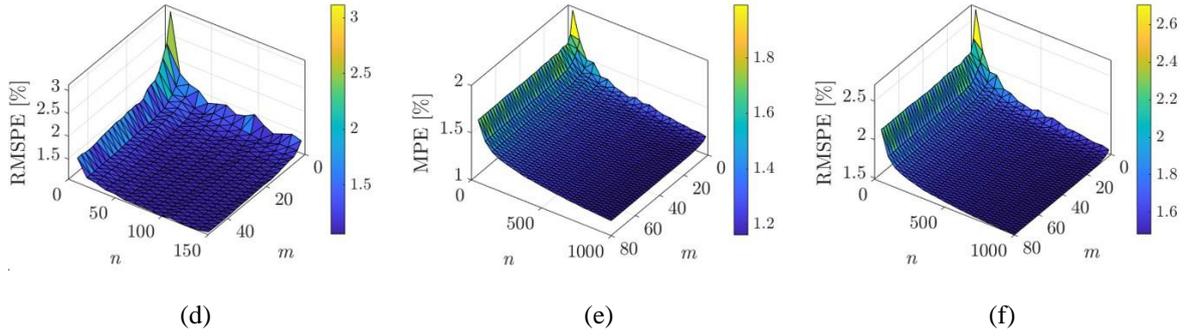

|  (d)  |  (e)  |  (f)  |

**Figure 5**: MPE$_{mean}$ and RMSPE$_{mean}$ as a function of different bag sizes and sample size for (a) NMC, (b) LCO, (c) LFP. The color bar corresponds to the error value. The plotted error is an average of 10 iterations.

With the LCO dataset, the average RMSPE and MPE error quickly decrease as the number of bags and sample size increase. The data needed to make better estimation was relatively more than the NMC dataset, which is to be expected due to the variable aging procedures. The average MPE is below 0.8% and RMSPE was close to 1% even with a relatively small $m$ and $n$ of 10 and 80, for example, with less memory usage compared to using all data (0.802 megabytes vs. 1.13 megabytes based on data from 14 cells). For LFP, using $n = 850$ and $m = 20$ results in error less than 1.5% in average RMSPE$_{mean}$ and 1.18% in average MPE$_{mean}$.

**Discussion**

Battery data is becoming more important and abundant than ever due to the rapid growth of the lithium-ion battery industry. This calls for an immediate need of efficient approaches to diagnose the SOH of batteries, especially for repurposing purposes. The proposed novel method consists of (a) using simple statistics to describe a small window of voltage and current data, (b) fast feature selection, and (c) bagging multiple GPR models.

Capacity estimation is made based on pure current and voltage charging data from datasets of diverse cathode chemistry undergoing different aging mechanisms. This work highlights the opportunity to develop data-driven based algorithms for quick and accurate estimation of capacity, which can find use in sorting retired batteries for second life applications despite their cycling history. The algorithm is an ensemble method containing GPR with bagging that outperforms conventional GPR based models both in terms of accuracy and computational time. It should be noted that the regions of voltage and current data used by the model to estimate the health can be adjusted. Further, the feature generation and ranking or selection process should work as presented, provided that within the selected region, changes in voltage and current are visible over time. Aging assessment of LCO and NMC dataset showed to be more accurate likely due to the relatively similar degradation over time between cells. Based on multiple training-test partitions, average and median RMSPE and MPE performance errors are found to be less than 1.48% and 1.29%, respectively. A consideration to be made also is the trade-off with uncertainty. Using similar data can cause overconfidence in models, while on the other hand, using not enough data will lead to loss in accuracy. The search space for $m$ and $n$, parameters of the bagged method, provides guidance on this trade-off since it is possible to easily observe the change in performance metrics.

Ultimately, the bagged approach can save on memory and processing power even with a rich amount of data. This can be expanded to practical applications due to the way the data is collected on the charging curve and serves as an effective diagnostic technique to evaluate feasibility of second-life applications. We acknowledge that there are other ways to scale Gaussian Processes in the literature, such as the Bayesian Committee Machine or approximating the kernel function, and these introduce interesting areas that can be explored in the context of battery degradation.[18,25,26]

**Limitation of study**

For future work, the scalability of data-driven models, in addition to accurate and safe health assessment is a key design parameter for transitioning models to industrial applications and achieving sustainability goals. Moreover, a chemistry-agnostic GPR with bagging algorithm can be designed as an extension of this work using richer data sets representative of real-world EV usage condition.[29]

**METHOD DETAILS**

**Datasets**

Datasets from online repositories were used in this work. Each of the datasets used unique aging experiments and different battery chemistries – namely, LFP, NMC, and LCO. Aging data used in this paper for all cells is shown in Fig. 1a, 1b, and 1c.

*Nickel Manganese Cobalt Oxide (NMC)*

Th NMC dataset[16] used 8 Kokam SLPB533459H4 NMC cells, which aged via the ARTEMIS driving schedule.[28] Characteristic data is taken for every 100 cycles with a 1C full charge-discharge cycle and a constant-current OCV test. The discharge capacity data is available hence it is used for SOH calculations.

*Lithium Cobalt Oxide (LCO)*

The LCO dataset uses statistically random discharge for battery deterioration.[17] The cells used were LG Chem. 18650 LCO cells. The cells were divided into multiple groups of 4, each undergoing a unique, randomized charging and discharging procedure at room temperature. A characteristic charge-discharge cycle at 2A took place periodically, allowing for a comparison between aging procedures. Capacity data is unavailable, so the discharge curve is integrated to obtain SOH measures.

*Lithium Iron Phosphate (LFP)*

In this group of cells, 124 commercial high-power LFP A123 APR18650M1A cells were aged via a full two-step fast charging cycle and 4C discharge.[5] Another batch of 45 cells from the same manufacturer underwent a 10 minute 4-step fast charge cycle and 4C discharge.[20] Once cells reached 80% nominal capacity a 1C charging regime followed by constant voltage (CV) charge was used to fully charge the cells in both datasets. There are three different batches of cells and each batch differing by the amount of rest taken between charging and discharging phases.[5] Discharge capacity data is available, so we use these measurements as the expected response.

**Data Preprocessing**

For data processing on MATLAB, erroneous measurements, such as battery voltage exceeding cutoff values, were first removed. An interpolation scheme is then used on the voltage and current charging curves to obtain regular measurements. The frequency of measurements used (measurements are needed every 2 – 10 seconds) for the ML model is slower than the frequencies used for lab data, which is often measured less than every second. Thus, a simple linear interpolation was enough for collecting accurate measurements of battery voltage and current. Measurements are then collected at the desired frequency based on the starting voltage or current value.

**Gaussian Process Regression Details**

The Matern 5/2 covariance function with automatic relevance determination is used for all three datasets since it is able to adapt to different smoothness throughout the regression problem.[12,23] The function is as follows:

$$k_f(x, x') = \sigma_f^2 \frac{2^{1-v}}{\Gamma(v)} \left(\sqrt{2v}\frac{x-x'}{\xi}\right)^v K_v\left(\sqrt{2v}\frac{x-x'}{\xi}\right), \tag{6}$$

where $x$ and $x'$ are any two points of the same variable, $\sigma_f^2$ is the signal variance of the input variables, $\Gamma$ is the gamma function, $K_v$ is the modified Bessel function, $\xi$ is the length scale, and $v = 5/2$ which simplifies the expression considerably. The hyperparameters that need to be optimized are the length scale, a measure that outlines how far extrapolations can be made, and the signal variance, which determines the spread of the joint distribution. Essentially, optimizing the set of hyperparameters allows the GPR model to fit a set of input variables $X$ to model a Gaussian with respect to the response $y$. The hyperparameters are optimized by maximizing the log marginal likelihood function, which allows for automatic tradeoff between bias and variance. The equation is as follows:

$$L = -\frac{1}{2}\log(\det(k_f + \sigma_n^2 I)) - \frac{1}{2}(y - H\beta)^T [K_f + \sigma_n^2 I]^{-1}(y - H\beta) \tag{7}$$
$$- \frac{N}{2}\log(2\pi),$$

where $k_f$ is the selected covariance function, $\sigma_n^2$ is the noise variance of the prediction, $I$ is the identity matrix, $H$ is the basis function (assumed to be 1), and $\beta$ is the coefficient for the basis function. For predictions of variables far from the training set, the state of health prediction will revert to $H\beta$. MATLAB's fitrgp function is used for optimizing the hyperparameters.

**Predictions with Boostrap Aggregating**

For bagging, the predictions of each model must be combined to generate a prediction. While weightless aggregation is possible, it has been shown that a weighted prediction is often better.[24] For this work, predicted responses are weighted based on the standard deviation. The weight function is as follows:

$$w_a = \frac{1}{\sigma_a}, \tag{8}$$

where $a$ is the $a$-th GPR model in the $m$ model set for a particular prediction, $\sigma$ is the associated error standard deviation. In general, this means that a more unconfident prediction is punished more, and it has been shown that a weighted average performs better.[27] The standard deviation is used instead of the variance since the variance more heavily favors or punishes individual models, which leads to significantly overconfident predictions with large $m$ and $n$.[26] This expression is multiplied with the corresponding prediction to establish a weighted average, which also has a weighted standard deviation as follows:

$$y_{pred} = \frac{\sum_{a=1}^{m} w_a y_a}{\sum_{a=1}^{m} w_a} \tag{9}$$

$$\sigma_{pred} = \sqrt{\frac{Z \sum_{a=1}^{m} w_a (y_a - y_{pred})^2}{(Z-1) \sum_{a=1}^{m} w_a}}, \tag{10}$$

where $y_{pred}$ is the aggregated SOH prediction, $y_a$ is the prediction made by the $a$-th model out of the $m$-model set, $\sigma_{pred}$ is the predicted standard deviation and $Z$ is the number of nonzero weights (which is almost always the same as $m$ in our implementation, but is included for greater generalizability).

**Model Performance Metrics**

Formally, RMSPE and MPE are defined as follows:

$$\text{RMSPE (\%)} = \sqrt{\frac{1}{c} * \sum_{j=1}^{c} \left(\frac{y_{pred,j}}{y_{exp,j}} - 1\right)^2} * 100\% \qquad (11)$$

$$\text{MPE (\%)} = \sum_{j=1}^{c} \left|\frac{y_{pred,j}}{y_{exp,j}} - 1\right| * \frac{100\%}{c}, \qquad (12)$$

where $j$ is the $j$-th datapoint in the $c$ characteristic cycles of a single cell, $y_{exp}$ is the expected SOH from the discharge data and $y_{pred}$ is the weighted average of the predicted SOH.